\documentclass[twocolumn]{aa}
\usepackage{graphicx}
\usepackage{txfonts}
\begin{document}
\title{Progenitor mass of the type IIP supernova 2005cs}

\author{V. P. Utrobin\inst{1,2} \and N. N. Chugai\inst{3}}

\offprints{V. Utrobin, \email{utrobin@itep.ru}}

\institute{
   Max-Planck-Institut f\"ur Astrophysik,
   Karl-Schwarzschild-Str. 1, D-85741 Garching, Germany
\and
   Institute of Theoretical and Experimental Physics,
   B.~Cheremushkinskaya St. 25, 117218 Moscow, Russia
\and
   Institute of Astronomy of Russian Academy of Sciences,
   Pyatnitskaya St. 48, 109017 Moscow, Russia}

\date{Received 28 May 2008 / accepted 29 August 2008}

\abstract{
The progenitor mass of type IIP supernova can be determined from either
   hydrodynamic modeling of the event or pre-explosion observations.
}{
To compare these approaches, we determine parameters of the sub-luminous
   supernova 2005cs and estimate its progenitor mass.
}{
We compute the hydrodynamic models of the supernova to describe its light
   curves and expansion velocity data.
}{
We estimate a presupernova mass of $17.3\pm1~M_{\sun}$, an explosion energy of
   $(4.1\pm0.3)\times10^{50}$ erg, a presupernova radius of
   $600\pm140~R_{\sun}$, and a radioactive $^{56}$Ni mass of
   $0.0082\pm0.0016~M_{\sun}$.
The derived progenitor mass of SN~2005cs is $18.2\pm1~M_{\sun}$, which is
   in-between those of low-luminosity and normal type IIP supernovae.
}{
The obtained progenitor mass of SN~2005cs is higher than derived
   from pre-explosion images.
The masses of four type IIP supernovae estimated by means of hydrodynamic modeling are
   systematically higher than the average progenitor mass for the
   $9-25~M_{\sun}$ mass range.
This result, if confirmed for a larger sample, would imply that a serious revision
   of the present-day view on the progenitors of type IIP supernovae is required.
}
%
\titlerunning{Progenitor of SN~2005cs}
\authorrunning{V. P. Utrobin \& N. N. Chugai}
\maketitle

\section{Introduction}
\label{sec:intro}
Type IIP supernovae (SNe~IIP) originate presumably from main-sequence stars
   of the mass range of $9-25~M_{\sun}$ (Heger et al. \cite{HFWLH_03}).
If this is the case, the predicted rate of SNe~IIP should follow the Salpeter
   initial mass function with half of events occurring for stars of mass below $13~M_{\sun}$.
This conjecture requires confirmation by means of the determination of progenitor
   mass for an extended sample of SNe IIP.

There are two ways to recover the progenitor mass of SN~IIP on the main sequence.
The first method is detection of the presupernova (pre-SN) in archival images
   of the host galaxy.
The estimated flux and color index of the detected pre-SN is then converted
   into a stellar mass using the flux and color index predicted by stellar
   evolution models.
Data for the available directly identified progenitors in the compilation of Li et al.
   (\cite{LWV_07}) indicate that for eight SNe IIP the progenitor masses have been
   estimated in this way, and for six SNe IIP, upper limits to the progenitor
   mass have been found.

An alternative approach to the mass determination involves hydrodynamic
   modeling of light curves and expansion velocities for the well-observed
   SNe IIP.
Combining the ejecta mass derived from the hydrodynamic modeling with the mass
   of the neutron star and the mass lost by the stellar wind provides us with
   the mass estimate of the progenitor.
Henceforth, the progenitor mass determined by this method is referred to as the
   "hydrodynamic mass".
At present, the hydrodynamic mass is measured only for three SNe IIP:
   the peculiar type IIP SN~1987A (Woosley \cite{W_88}; Blinnikov et al.
   \cite{BLBNI_00}; Utrobin \cite{U_05}), the normal type IIP SN~1999em
   (Baklanov et al. \cite{BBP_05}; Utrobin \cite{U_07}), and the
   low-luminosity type IIP SN~2003Z (Utrobin et al. \cite{UCP_07}).
The small amount of SN IIP events with the measured hydrodynamic mass is
   related to the fact that the hydrodynamic modeling requires a complete
   multi-band photometry at both the plateau and the radioactive tail, and spectra
   of sufficient quality.

There are only a few other SNe IIP that also meet these requirements.
Among these is the sub-luminous type IIP SN~2005cs (Pastorello et al.
   \cite{PST_06}).
On the basis of its luminosity and expansion velocities, this SN is intermediate between
   the low-luminosity and normal SNe~IIP.
Parameters of SN~2005cs are of considerable interest for two major reasons:
   (1) this SN is expected to have intermediate parameters, which would be
   interesting to check;
   (2) the pre-SN was detected in the pre-explosion images and its progenitor
   mass was estimated by several groups (Maund et al. \cite{MSD_05};
   Li et al. \cite{LVF_06}; Eldridge et al. \cite{EMS_07}) providing
   an opportunity to compare different mass estimates.

In the paper, we perform hydrodynamic modeling of SN 2005cs to recover the
   parameters: ejecta mass, explosion energy,
   pre-SN radius, and radioactive $^{56}$Ni mass.
We start with the description of the model and observational data used
   and then present the results of the hydrodynamic modeling for SN~2005cs
   (Sect.~\ref{sec:opthydmod}).
In Sect.~\ref{sec:why}, we present additional arguments in favor of our choice
   of pre-SN models.
Possible uncertainties in the hydrodynamic mass of SN 2005cs are analyzed
   (Sect.~\ref{sec:prog}), and finally the implications of hydrodynamically
   studied objects for the origin of SNe~IIP are discussed
   (Sect.~\ref{sec:discon}).

We adopt an explosion date on June 27.5 UT (JD 2453549) and a distance of 8.4 Mpc
   following Pastorello et al. (\cite{PST_06}), and a reddening $E(B-V)=0.12$
   taken from Li et al. (\cite{LVF_06}).

\section{Optimal hydrodynamic model}
\label{sec:opthydmod}
%
\begin{figure}[b]
   \resizebox{\hsize}{!}{\includegraphics{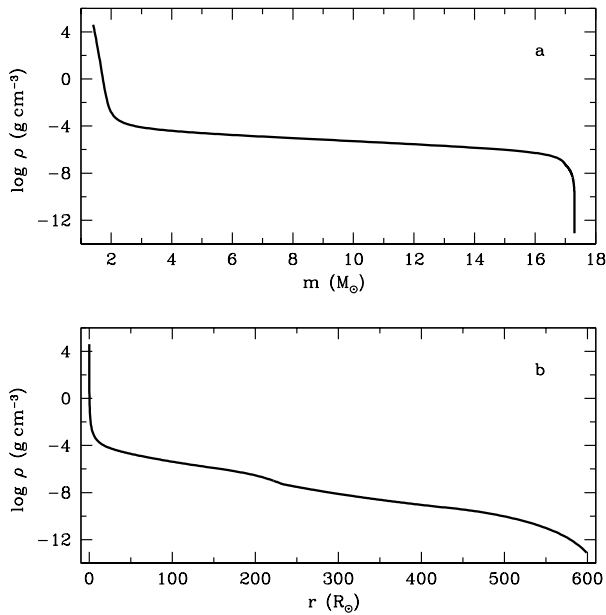}}
   \caption{%
   Density distribution as a function of interior mass \textbf{a}) and
   radius \textbf{b}) for the optimal pre-SN model.
   The central core of 1.4 $M_{\sun}$ is omitted.
   }
   \label{fig:denmr}
\end{figure}
The hydrodynamic model applied to SN~2005cs is essentially the same as used
   before for SN~1999em (Utrobin \cite{U_07}).
The pre-SN structure is set to be a non-evolutionary model of a red supergiant
   (RSG) star (but see Sect.~\ref{sec:why}).
The chemical composition of the hydrogen envelope is solar.
Although this might be a simplification, there is no observational
   evidence that the hydrogen abundance in the atmosphere of the RSG, e.g.
   $\alpha$~Ori, differs notably from solar (Harper et al. \cite{HBL_01}).
The effect of the variation in the surface abundances on the SN~IIP light curve
   was studied before, and it was found that even a significant
   change in the hydrogen abundance of outer layers only weakly affects
   the light curve (Utrobin \cite{U_07}).
The light curve of the SN~2005cs model of hydrogen abundance X=0.65 and
   helium abundance Y=0.33, predicted for an $18~M_{\sun}$ star with
   initial solar composition (Heger \cite{Heg_98}), is almost
   indistinguishable from the case of solar composition.
The inner layers of the ejecta are mixed with the helium core in the same way
   as in the model for SN~1999em.

We refer to the hydrodynamic model as being the optimal one, in terms of an
   "eye-fit" to the observational light curve and the evolution of photospheric
   velocity.
In general, a numerical optimization procedure could be developed to complete
   the search for the best-fit model.
However, at present it would require an enormous amount of computational time
   that would be unjustified because the error of eye-fit is far less
   than the error introduced by uncertainties in the distance and
   interstellar extinction.
The search for the optimal model of SN~2005cs uses the hydrodynamic
   model behavior in parameter space studied earlier in detail
   (Utrobin \cite{U_07}).
The $^{56}$Ni mass is determined empirically from the comparison of the
   $R$-band luminosity of SN 2005cs at the radioactive tail with that of
   SN 1987A.
For the adopted distance and reddening, the $R$-values at the age of 250--300
   days (Tsvetkov et al. \cite{TVS_06}) correspond to the $^{56}$Ni mass of
   $0.0082~M_{\sun}$.

The observed bolometric light curve of SN~2005cs is recovered from $UBVRI$
   photometry (Pastorello et al. \cite{PST_06}; Tsvetkov et al. \cite{TVS_06})
   using a black-body approximation for the SN radiation, while expansion
   velocities at the photosphere were taken from Pastorello et al. (\cite{PST_06}).

\begin{figure}[t]
   \resizebox{\hsize}{!}{\includegraphics[clip, trim=0 0 0 184]{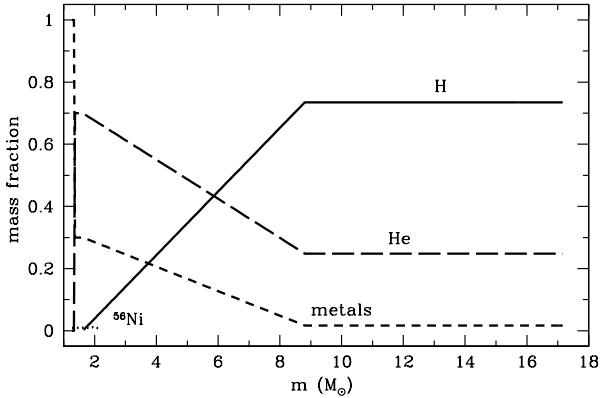}}
   \caption{%
   The mass fraction of hydrogen (\emph{solid line\/}), helium
      (\emph{long dashed line\/}), heavy elements (\emph{short dashed line\/}),
      and radioactive $^{56}$Ni (\emph{dotted line\/}) in the ejecta of
      the optimal model.
   }
   \label{fig:chcom}
\end{figure}
\begin{figure}[b]
   \resizebox{\hsize}{!}{\includegraphics{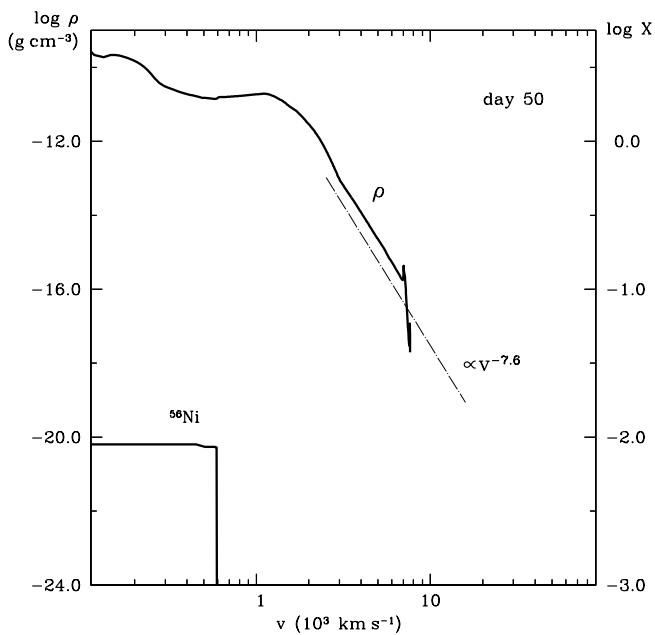}}
   \caption{%
   The density and the $^{56}$Ni mass fraction as a function of the velocity
      for the optimal model at $t=50$ days.
   \emph{Dash-dotted line} is the density distribution fit
      $\rho \propto v^{-7.6}$.
   }
   \label{fig:denicl}
\end{figure}
\begin{figure}[t]
   \resizebox{\hsize}{!}{\includegraphics[clip, trim=0 0 0 184]{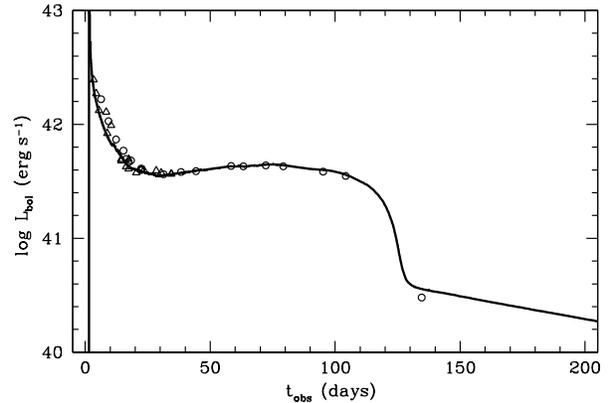}}
   \caption{%
   Comparison of the calculated bolometric light curve of the optimal model
      (\emph{solid line\/}) with the bolometric data of SN 2005cs
      evaluated from the photometric observations of Pastorello et al.
      (\cite{PST_06}) (\emph{open triangles\/}) and Tsvetkov et al.
      (\cite{TVS_06}) (\emph{open circles\/}).
   }
   \label{fig:lmbol}
\end{figure}
The dependence of the light curve at the initial adiabatic cooling stage and
   the plateau phase on SN parameters provides us with a toolkit to
   search for the optimal model in parameter space.
The obtained model of SN 2005cs is characterized by the ejecta mass
   $M_{env}=15.9~M_{\sun}$, the explosion energy $E=4.1\times10^{50}$ erg,
   and the pre-SN radius $R_0=600~R_{\sun}$ with the $^{56}$Ni mass
   $M_{\mathrm{Ni}}=0.0082~M_{\sun}$.
The ejecta mass combined with the neutron star mass results in the pre-SN mass
   of $17.3~M_{\sun}$.

The optimal pre-SN density structure is shown in Fig.~\ref{fig:denmr} and
   the chemical composition in Fig.~\ref{fig:chcom}.
The helium core is mixed with the hydrogen envelope so that the hydrogen
   abundance increases linearly with mass in the inner $9~M_{\sun}$.
We adopt the helium-core mass of $5.4~M_{\sun}$, which corresponds to the
   $\approx 18~M_{\sun}$ progenitor (Hirschi et al. \cite{HMM_04}).
We note that the basic SN parameters are insensitive to the helium-core mass
   (Utrobin et al. \cite{UCP_07}).
In the freely expanding envelope, the hydrogen is mixed downward to
   300 km\,s$^{-1}$, while the radioactive $^{56}$Ni is mixed outward to
   610 km\,s$^{-1}$ (Fig.~\ref{fig:denicl}).

The observed bolometric light curve is reproduced by our optimal model
   (Fig.~\ref{fig:lmbol}).
A small disparity between model and observations at the radioactive tail is
   probably caused by the black-body approximation applied to the reconstruction
   of the observed bolometric light curve.
This approximation is certainly rough at the nebular epoch.
We note that the $^{56}$Ni mass was derived from the comparison of the $R$-band
   luminosities of SN~2005cs and SN~1987A at similar nebular epochs.

\begin{figure*}[t]
   \centering
   \includegraphics[width=17cm]{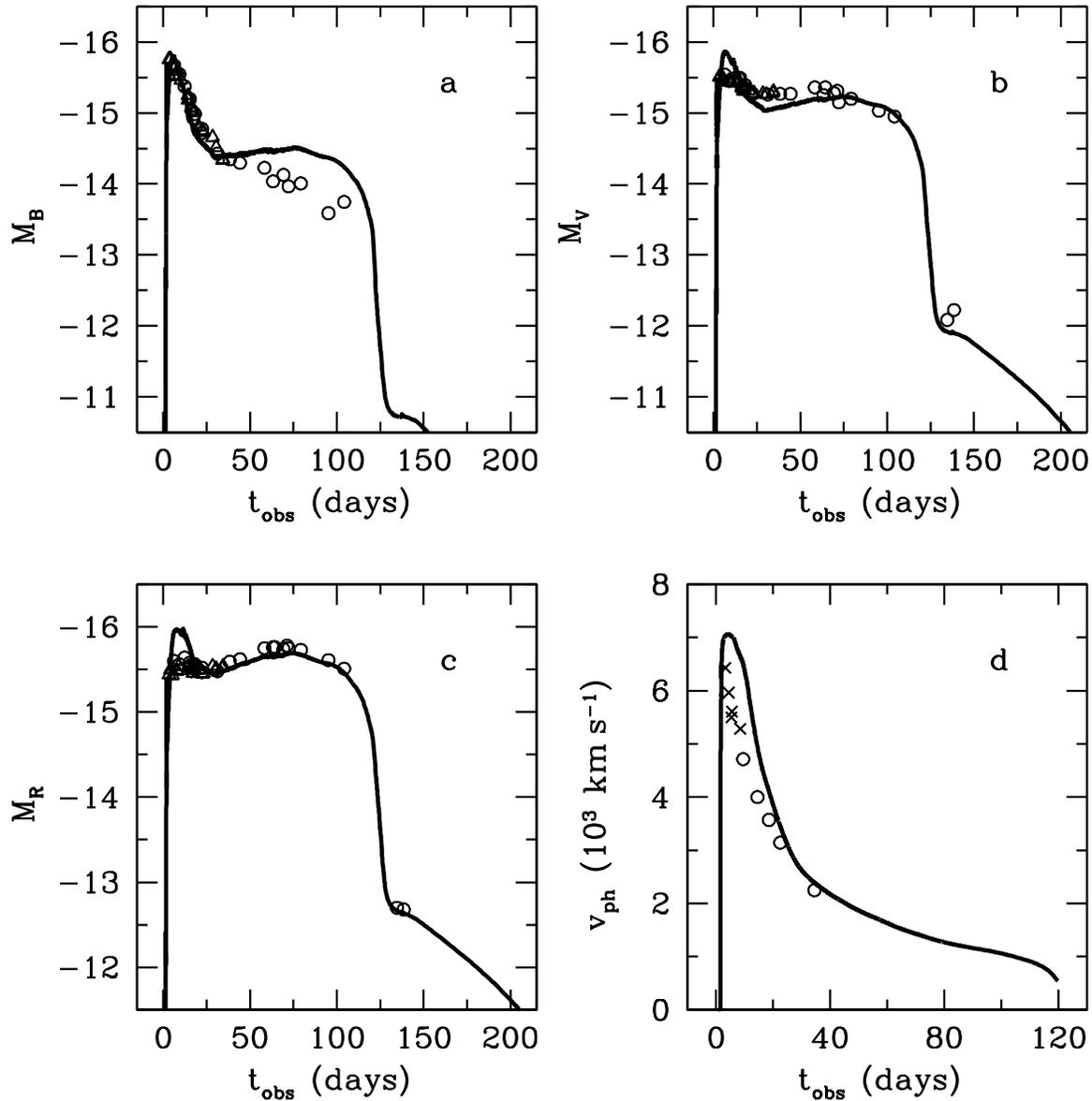}
   \caption{%
   Optimal hydrodynamic model.
   Panel \textbf{a}): the calculated $B$ light curve (\emph{solid line\/})
      compared with the observations of SN 2005cs obtained by Pastorello et al.
      (\cite{PST_06}) (\emph{open triangles\/}) and Tsvetkov et al.
      (\cite{TVS_06}) (\emph{open circles\/}).
   Panels \textbf{b}) and \textbf{c}): the same as panel \textbf{a}) but for
      the $V$ and $R$ light curves.
   Panel \textbf{d}): calculated photospheric velocity (\emph{solid
      line\/}) is compared with photospheric velocities estimated from
      absorption minima of the He\,I 5876 \AA\ line (\emph{crosses\/})
      and the Fe\,II 5169 \AA\ line (\emph{open circles\/}) measured by
      Pastorello et al. (\cite{PST_06}).
   }
   \label{fig:optmod}
\end{figure*}
Although the one-group approximation for radiation transfer prevents us from
   achieving a detailed description of the spectral energy distribution, our model does
   reproduce the general magnitude of the broad-band photometry, although the match
   is not precise.
Both the observed and calculated $B$ light curves (Fig.~\ref{fig:optmod}a)
   show an initial peak related to the cooling of hot outer layers of shocked
   ejecta.
The amplitude and width of this peak is sensitive to the structure of
   the outermost rarefied layers of the pre-SN envelope, and the density
   structure shown in Fig.~\ref{fig:denmr} is optimal in this sense.
The later ($t>50$ days) behavior of the flux in $B$ band is poorly reproduced
   because the black-body approximation of the model spectrum in the
   $\lambda<4500$ \AA\ range is too crude at the plateau stage.
Fortunately, this discrepancy does not affect the model fit to the bolometric
   light curve because of a small contribution of the $B$ band to the
   bolometric luminosity at this stage.
The calculated $V$ and $R$ light curves for this model describe satisfactorily
   observations (Figs.~\ref{fig:optmod}b and c).
The computed evolution of the photospheric velocity is also consistent with
   observations (Fig.~\ref{fig:optmod}d), supporting the detailed
   hydrodynamic properties of the model ejecta, in particular, the density
   distribution in the SN envelope (Fig.~\ref{fig:denicl}).

To estimate a measure of uncertainty in the derived physical
   parameters, we investigate the sensitivity of the optimal model to
   observational values.
We adopt the following relative changes in the observational values: 20\% in
   the bolometric luminosity, 5\% in the photospheric velocity, and 5\% in
   the plateau duration.
Using the auxiliary hydrodynamic models in the vicinity of the optimal model,
   we transform the adopted changes into changes in the pre-SN radius
   of $\pm140~R_{\sun}$, the ejecta mass of $\pm1~M_{\sun}$, the explosion
   energy of $\pm0.3\times10^{50}$ erg, and the $^{56}$Ni mass of
   $\pm0.0016~M_{\sun}$.
Given the errors in the observational values, these relations can be used to
   derive the errors in the physical parameters.
The adopted relative changes in the observed values are close to their typical
   errors.
We therefore consider the derived changes in the physical parameters of SN 2005cs
   to represent the typical uncertainties in these values.

\section{Why non-evolutionary presupernova model?}
\label{sec:why}
%
\begin{table}[t]
\caption[]{Hydrodynamic models for evolutionary presupernovae.}
\label{tab:evhydm}
\centering
\begin{tabular}{@{ } c @{ } c @{ } c @{ } c @{ } c @{ } c @{ } c @{ } c @{ } c @{ }}
\hline\hline
\noalign{\smallskip}
 Model & $R_0$ & $M_{env}$ & $E$ & $M_{\mathrm{Ni}}$ & $M_{\mathrm{He}}^{core}$
       & $M_{pre-SN}$ & $X$ & $Y$ \\
       & $(R_{\sun})$ & $(M_{\sun})$ & ($10^{50}$ erg) & $(10^{-2} M_{\sun})$
       & $(M_{\sun})$ & $(M_{\sun})$ & & \\
\noalign{\smallskip}
\hline
\noalign{\smallskip}
 EM1 & 1200 & 15.9 & 4.1 & 0.82 & 5.4 & 17.3 & 0.65 & 0.33 \\
 EM2 &  600 & 15.9 & 4.1 & 0.82 & 5.4 & 17.3 & 0.65 & 0.33 \\
 EM3 &  800 & 10.6 & 1.5 & 0.82 & 3.0 & 12.0 & 0.65 & 0.33 \\
 EM4 &  700 &  7.8 & 1.4 & 0.82 & 2.0 &  9.0 & 0.65 & 0.33 \\
\noalign{\smallskip}
\hline
\end{tabular}
\end{table}
Although our non-evolutionary pre-SN model closely resembles the massive RSG
   star by the heterogeneous core-envelope structure, the extended radius, and
   the helium-core mass, it omits a sharp jump in density and chemical
   composition between the helium core and hydrogen envelope, which is
   characteristic of the evolutionary model.
A question may then arise: why not use evolutionary pre-SN.
The answer is that the assumption of a smooth transition from the helium core
   to the hydrogen envelope in the non-evolutionary pre-SN is dictated by
   two major facts.
First, the explosion of the evolutionary model generally fails to reproduce the
   light curve of SN~IIP in detail, as became clear after SN~1987A
   (cf. Woosley \cite{W_88}).
Second, the H$\alpha$ profile in the SN 1987A spectra provides clear evidence
   of hydrogen mixing deep down inside the helium core.
We note that in 2D simulations the shock propagation produces
   Rayleigh-Taylor (RT) mixing between the oxygen and helium-core material, and
   between the helium-core matter and the hydrogen envelope
   (Herant \& Benz \cite{HB_91}; M\"uller et al. \cite{MFA_91};
   Kifonidis et al. \cite{KPSJM_03}, \cite{KPSJM_06}).

In general, a self-consistent hydrodynamic model of the explosion of the
   evolutionary RSG star should be three-dimensional in considering
   both the hydrodynamic flow and radiation transfer.
Unfortunately, this approach cannot presently be realized.
We therefore accounted for 3D effects in our 1D simulations by adopting
   a non-evolutionary pre-SN with density and chemical composition jumps that had been
   smoothed presumably by the RT mixing between the helium core and the hydrogen
   envelope.
This approach was justified because the RT mixing occurs before the
   shock breakout and does not affect the light curve directly.

\begin{figure}[t]
   \resizebox{\hsize}{!}{\includegraphics{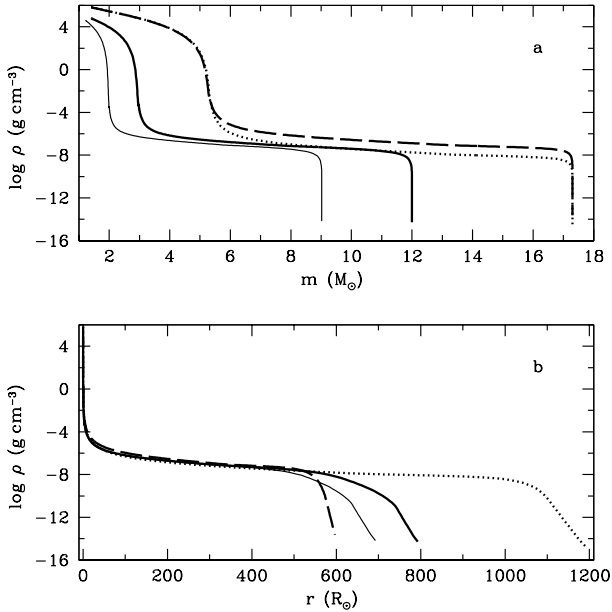}}
   \caption{%
   Density distribution as a function of interior mass \textbf{a}) and
      radius \textbf{b}) for the evolutionary pre-SNe (Table~\ref{tab:evhydm}):
      model EM1 (\emph{dotted line\/}), model EM2 (\emph{dashed line\/}),
      model EM3 (\emph{thick solid line\/}), and
      model EM4 (\emph{thin solid line\/}).
   }
   \label{fig:evmdn}
\end{figure}
Here we propose that a pre-SN has an evolutionary density structure and
   chemical composition.
We constructed the hydrostatic configuration, which reproduced the main features
   of evolutionary pre-SN models neglected before, namely the dense
   helium core with a sharp density gradient and the chemical composition
   jump at its boundary.
We refer to this pre-SN model as the "evolutionary model".
Four representative models with a dense helium core and extended hydrogen
   envelope (Fig.~\ref{fig:evmdn}) were considered.
Their parameters are listed
   in Table~\ref{tab:evhydm}, i.e. the pre-SN radius, ejecta mass,
   explosion energy, total $^{56}$Ni mass, helium-core mass,
   pre-SN mass, and surface hydrogen and helium abundances.
The helium-core masses are typical of massive RSG stars
   (Hirschi et al. \cite{HMM_04}; Garcia-Berro et al. \cite{GRI_97}).
The first two hydrodynamic models EM1 and EM2 are similar to the optimal model
   in the basic parameters but model EM1 differs in the initial radius.
The remaining two models EM3 and EM4 are less massive than the optimal model and
   their masses are close to the progenitor mass of SN~2005cs estimated from
   the pre-explosion images of the galaxy M~51 (see Sect.~\ref{sec:prog}).
We consider also the mixed models EM3 and EM4 in which the helium core
   is mixed by the hydrogen envelope --- in the same way as the optimal model
   is mixed (Fig.~\ref{fig:chcom}) --- in the inner $4.7~M_{\sun}$ and
   $2.8~M_{\sun}$, respectively.

\begin{figure*}[t]
   \centering
   \includegraphics[width=17cm]{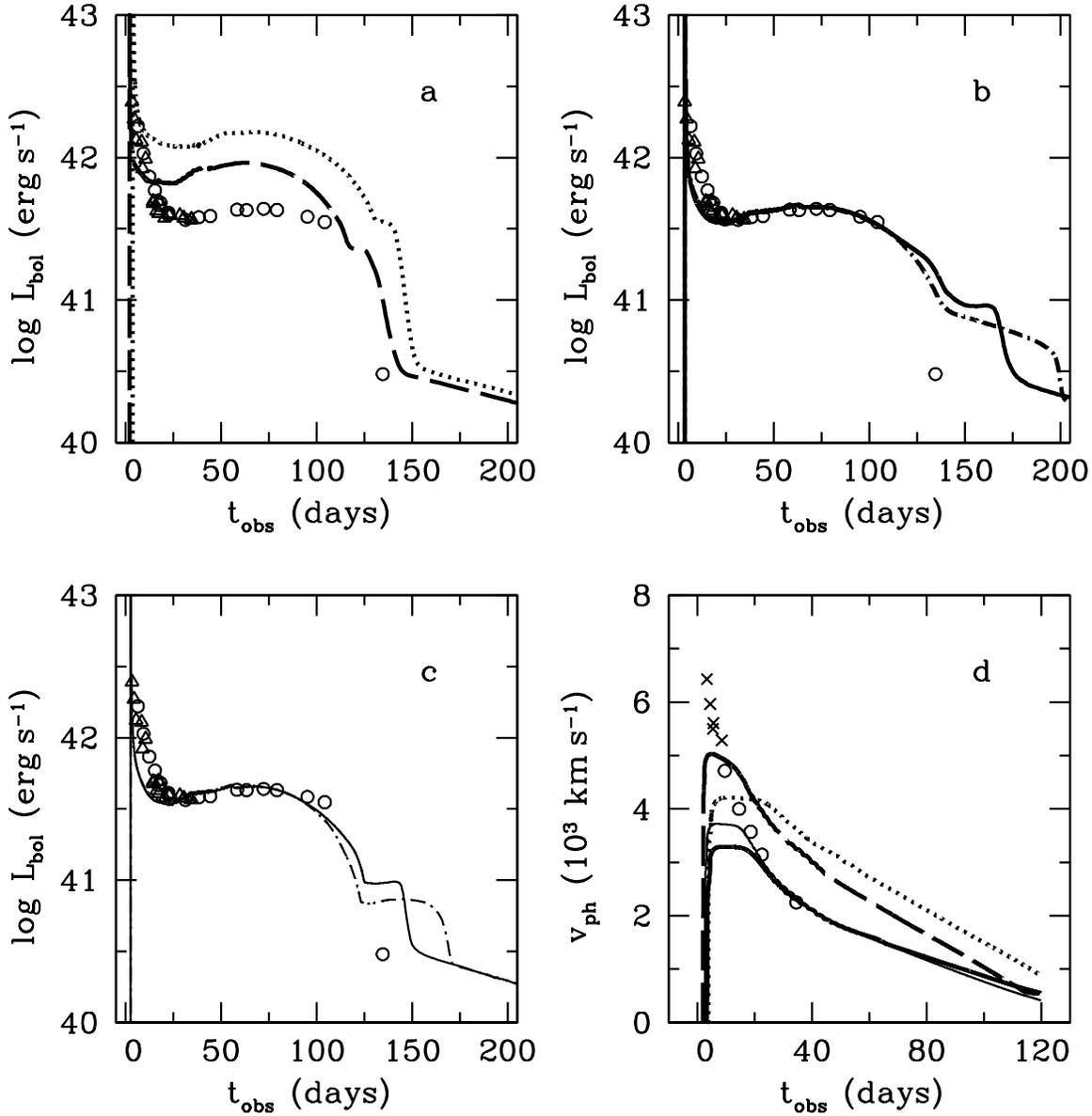}
   \caption{%
   Bolometric light curve \textbf{a}), \textbf{b}), and \textbf{c}) and
      photospheric velocity \textbf{d}) of the hydrodynamic models
      in Table~\ref{tab:evhydm} compared with the empirical data of SN 2005cs
      (see legends of Figs.~\ref{fig:lmbol} and \ref{fig:optmod} for details).
   \emph{Dotted line\/} is model EM1, \emph{dashed line\/} is model EM2,
      \emph{thick solid line\/} is model EM3, and \emph{thin solid line\/} is
      model EM4.
   Mixed models EM3 and EM4 are shown by the corresponding \emph{dotted-dashed
      lines\/} in panels \textbf{b}) and \textbf{c}).
   }
   \label{fig:evmlc}
\end{figure*}
First of all, we see that the explosion of the evolutionary model produces
   the dome-shaped light curve without a steep transition to the radioactive
   tail (Figs.~\ref{fig:evmlc}a, b, and c).
We note that the steep decline in luminosity at the end of the plateau is
   a characteristic of all SNe~IIP.
The dome shape of the light curve for the evolutionary model is
   related to the almost flat density distribution in the hydrogen envelope
   (Fig.~\ref{fig:evmdn}).
A step-like bump in the light curve at the transition to the radioactive tail
   is caused by the dense helium core (Fig.~\ref{fig:evmdn}).
We note that mixing in models EM3 and EM4 does not remove but modifies the
   step-like feature (Figs.~\ref{fig:evmlc}b and c).
The fact that this bump is never observed in SNe~IIP indicates that the density
   jump between the dense helium core and flat hydrogen envelope is smoothed
   and, consequently, the inner SN ejecta is strongly mixed.

The hydrodynamic model EM1 with an initial radius of $1200~R_{\sun}$,
   close to that of the evolutionary pre-SN (Heger \cite{Heg_98}), predicts an
   unacceptably high luminosity, long plateau (Fig.~\ref{fig:evmlc}a), and low
   photospheric velocity during the early epoch $t<10$ days
   (Fig.~\ref{fig:evmlc}d).
Model EM2 with the smaller initial radius, as expected, is characterized by
   lower luminosity, but remains too luminous (Fig.~\ref{fig:evmlc}a)
   and has low photospheric velocity at the early epoch $t<7$ days
   (Fig.~\ref{fig:evmlc}d).
The reduction in radius would produce a reasonable luminosity at the early plateau
   stage, but this model would have too low luminosity at the late plateau
   stage.
We therefore conclude that the replacement of the non-evolutionary pre-SN in the
   optimal model by the evolutionary pre-SN cannot produce a reasonable fit to
   the observational data.

Models EM3 and EM4 of lower mass and explosion energy could reproduce
   the observed plateau luminosity for an appropriate choice of initial
   radius (Figs.~\ref{fig:evmlc}b and c).
However, a full description of the light curve cannot be attained for either
   the $12~M_{\sun}$ or $9~M_{\sun}$ pre-SNe.
The photospheric velocity at the early stage in these models becomes
   unacceptably low (Fig.~\ref{fig:evmlc}d), which is another reason why
   evolutionary models have to be modified significantly to attain a reasonable fit
   to observational data.
Remarkably, all evolutionary models fail to reproduce the width of the
   initial peak.
We therefore conclude that there is no set of the basic parameters for the
   $9-18~M_{\sun}$ progenitors that can reproduce the observations of SN 2005cs in
   the frame of the 1D explosion of the evolutionary pre-SNe.

\section{Progenitor mass}
\label{sec:prog}
To derive the main-sequence mass of the SN 2005cs progenitor, the pre-SN mass
   needs to be combined with the mass lost during the hydrogen and helium burning
   stages.
The mass lost at the hydrogen burning stage is taken to be $\approx0.25~M_{\sun}$,
   the average value between the masses lost in the $15~M_{\sun}$ and
   $20~M_{\sun}$ evolutionary models of non-rotating stars developed by
   Meynet et al. (\cite{MM_03}), who assumed the theoretical mass-loss rate
   provided by Vink et al. (\cite{VKL_01}).

The mass lost at the helium burning stage can be estimated from the stage duration,
   which is a function of the initial mass $M$ and mass-loss rate
   $\dot{M}$ adopted for the RSG stage.
We adopt the duration of the helium burning stage from Meynet et al. (\cite{MM_03}).
The dependence of the mass-loss rate $\dot{M}$ on $M$ was taken from Chevalier
   et al. (\cite{CFN_06}), who used $\dot{M}$ values from de Jager et al.
   (\cite{JNH_88}).
We also used the calibration $\dot{M}=1.5\times10^{-6}~M_{\sun}$\,yr$^{-1}$ for
   the RSG with a main-sequence star $M=22~M_{\sun}$, on the basis of the
   SN~1999em study (Chugai et al. \cite{CCU_07}).
The derived mass lost at the RSG stage was $0.6~M_{\sun}$.
The total mass lost by the stellar wind was $\approx0.85~M_{\sun}$,
   and the main-sequence mass of the progenitor was $18.2~M_{\sun}$.

The mass of the SN~2005cs progenitor was estimated
   from archival images of the galaxy M51 taken by the Advanced Camera for
   Surveys of the \emph{Hubble Space Telescope} (\emph{HST\/}) and from
   near-infrared images acquired by the Near Infrared Camera and Multi-Object
   Spectrometer on board \emph{HST\/} in $JHK$ bands.
We note that the progenitor was detected only in the $I$\/-band image; in other
   bands, only upper limits to fluxes were obtained.
From these data, Maund et al. (\cite{MSD_05}) derived a progenitor mass of
   $9^{+3}_{-2}~M_{\sun}$, Li et al. (\cite{LVF_06}) reported a progenitor
   mass of $10\pm3~M_{\sun}$, while Eldridge et al. (\cite{EMS_07}) derived
   a progenitor mass of between $6~M_{\sun}$ and $8~M_{\sun}$.
These estimates therefore propose a $6-13~M_{\sun}$ range for the progenitor mass of
   SN~2005cs.

The above mass estimates are significantly lower than our hydrodynamic mass.
The disagreement is serious and requires an explanation.
Our hydrodynamic model for SN~IIP (Utrobin \cite{U_04}) can be checked by
   comparison with the independent model of Blinnikov et al.
   (\cite{BEBPW_98}).
In the case of the normal type IIP SN~1999em, both codes produce
   similar ejecta mass by assuming the same distance (Baklanov et al.
   \cite{BBP_05}; Utrobin \cite{U_07}).
We explored crucial model assumptions that might minimize the ejecta mass.
One critical point is the degree of mixing between the hydrogen
   envelope and helium core.
We found that the minimal mass was produced, if complete mixing occurred.
In this case, the ejecta mass of the hydrodynamic model could be reduced by about
   $0.5~M_{\sun}$.
Another uncertainty was related to the incompleteness of the line list
   used in the line-opacity calculations.
By studying this issue using the latest line list of Kurucz with
   $\approx 6.2\times10^7$ observed and predicted lines, it was found to provide
   only negligible effect, which may cause the mass decrease by the value of
   the order of $0.1~M_{\sun}$.
Both uncertainties implied a lower limit for the hydrodynamic progenitor
   mass as low as $17.6~M_{\sun}$, which is higher than the upper limit
   of $13~M_{\sun}$ recovered from the pre-explosion images of SN~2005cs.

\section{Discussion and conclusions}
\label{sec:discon}
%
\begin{table}[t]
\caption[]{Hydrodynamic models for SN 1987A, SN 1999em, SN 2003Z, and SN 2005cs.}
\label{tab:sumtab}
\centering
\begin{tabular}{@{ } l @{ } c @{ } c @{ } c @{ } c @{ } c @{ } c @{ } c @{ }}
\hline\hline
\noalign{\smallskip}
 SN & $R_0$ & $M_{env}$ & $E$ & $M_{\mathrm{Ni}}$ & $Z$
       & $v_{\mathrm{Ni}}^{max}$ & $v_{\mathrm{H}}^{min}$ \\
       & $(R_{\sun})$ & $(M_{\sun})$ & ($10^{51}$ erg) & $(10^{-2} M_{\sun})$
       & & (km\,s$^{-1}$) & (km\,s$^{-1}$) \\
\noalign{\smallskip}
\hline
\noalign{\smallskip}
 87A &  35 & 18   & 1.5   & 7.65 & 0.006 & 3000 & 600 \\
99em & 500 & 19   & 1.3   & 3.60 & 0.017 &  660 & 700 \\
 03Z & 229 & 14   & 0.245 & 0.63 & 0.017 &  535 & 360 \\
05cs & 600 & 15.9 & 0.41  & 0.82 & 0.017 &  610 & 300 \\
\noalign{\smallskip}
\hline
\end{tabular}
\end{table}
\begin{figure}[t]
   \resizebox{\hsize}{!}{\includegraphics{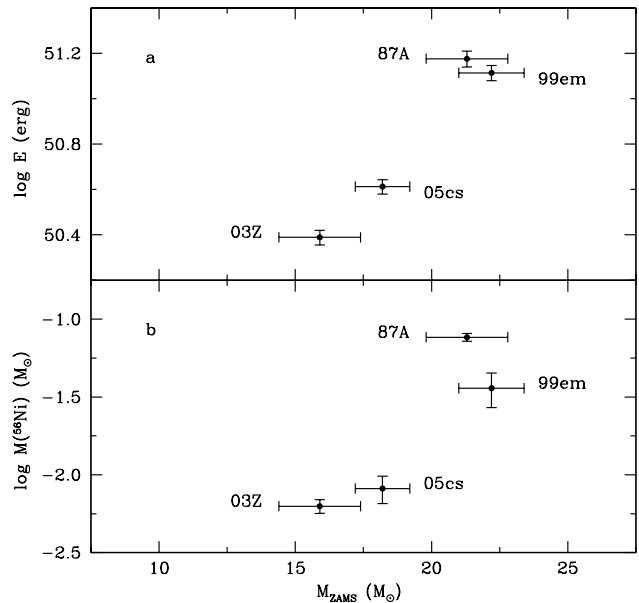}}
   \caption{%
   Explosion energy \textbf{a}) and $^{56}$Ni mass \textbf{b}) versus
      hydrodynamic progenitor mass for four core-collapse SNe.
   }
   \label{fig:nienms}
\end{figure}
The primary goal of this study was to determine parameters of the sub-luminous
   type IIP SN 2005cs by means of hydrodynamic modeling.
We have estimated a pre-SN mass of $17.3\pm1~M_{\sun}$, explosion energy of
   $(4.1\pm0.3)\times10^{50}$ erg, pre-SN radius of $600\pm100~R_{\sun}$,
   and $^{56}$Ni mass of $0.0082\pm0.0016~M_{\sun}$.
Using conservative assumptions about the mass-loss rate at the hydrogen and
   helium burning stages, we estimate the main-sequence progenitor mass of
   $17.2-19.2~M_{\sun}$.

Hydrodynamic models for four SNe~IIP listed in Table~\ref{tab:sumtab} are
   characterized by the pre-SN radius, the ejecta mass, the explosion energy,
   the total $^{56}$Ni mass, the surface abundance of heavy elements,
   the maximum velocity of nickel, and the minimum velocity of the hydrogen-rich
   envelope.
The major parameters of SN 2005cs --- the ejecta mass, the explosion energy,
   and the $^{56}$Ni mass --- are intermediate
   between those of the low-luminosity type IIP SN 2003Z
   (Utrobin et al. \cite{UCP_07}) and the normal type IIP SN 1999em
   (Utrobin \cite{U_07}) in qualitative agreement with their luminosities.
At present, there are therefore four SNe~IIP that have parameters determined
   by hydrodynamic modeling.
For these objects, the explosion energy and $^{56}$Ni mass correlate with
   the progenitor mass (Fig.~\ref{fig:nienms}).
This is consistent with the empirical relation between the explosion energy and
   $^{56}$Ni mass found by Nadyozhin (\cite{N_03}) for normal SNe~IIP.

Despite the uncertainties in hydrodynamic modeling, the disparity between the
   hydrodynamic mass of the SN~2005cs progenitor and the mass estimated from
   the pre-explosion images is significant.
This difference could be decreased by including the effects of the pre-SN light
   absorption in a hypothetical dusty circumstellar shell.
However, this issue requires careful consideration, which is beyond the scope of
   our paper.
We note only that this conjecture has a number of observational implications
   need to be to verified.
The presence of the dense dusty shell around pre-SN should produce
   strong Na\,I absorption lines in the SN~IIP spectrum at the photospheric
   epoch.
In the case of a normal SN~IIP and normal pre-SN wind without a dense
   circumstellar shell, the predicted Na\,I absorptions are weak and
   probably not observable (Chugai \& Utrobin \cite{CU_08}).
In addition, the interaction of the SN ejecta with the dense circumstellar shell
   should produce an outburst of radio and X-ray emission at an age of
   about $\sim10^2$ days.
The most apparent effect of the light absorption in the dusty circumstellar
   shell should be a large $J-K$ color index of the pre-SN.
For instance, we have found that the pre-SN light absorption, required
   to allow massive progenitor implied by both the pre-explosion $I$ value
   and upper limits in other bands for SN~2005cs, suggests a large color
   index $J-K\approx2.5$ mag, compared with the intrinsic $J-K$ index of $0.7-1$ mag
   for typical galactic K-M supergiants (Elias et al. \cite{EFH_85}).
In this regard, it is noteworthy that the type IIP SN~2008bk in the nearby
   galaxy NGC 7793 with available $JK$ photometry of the progenitor was found to have
   a moderate color index $J-K\approx1$ mag (Maoz \& Mannucci \cite{MM_08})
   which indicates little (if any) absorption.
The observational and hydrodynamic studies of this supernova would be of significant
   importance in clarifying the serious and challenging problem of
   the progenitor mass of SN~2005cs and in general SNe~IIP.

\begin{figure}[t]
   \resizebox{\hsize}{!}{\includegraphics[clip, trim=0 0 0 184]{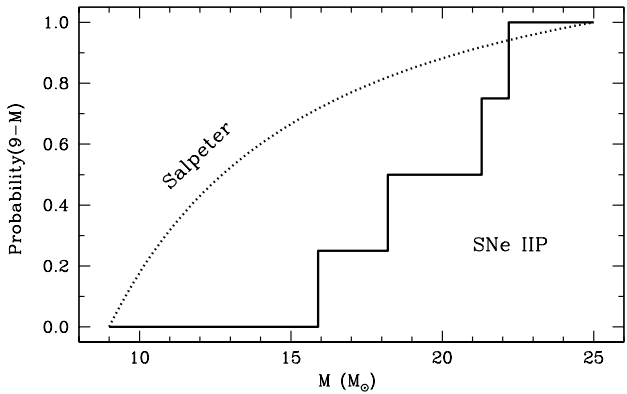}}
   \caption{%
   Cumulative "Salpeter" distribution of SN~IIP progenitors in
      the $9-25~M_{\sun}$ mass range and the distribution of
      hydrodynamic progenitor masses of four SNe~IIP.
   }
   \label{fig:cumdis}
\end{figure}
The number of SNe~IIP with measured hydrodynamic masses is too small to
   be able to analyze in detail and draw reliable conclusions about their
   mass distribution.
However, the hydrodynamic progenitor masses do appear
   to be systematically higher than if SNe~IIP had originated
   from the range of $9-25~M_{\sun}$, assuming a Salpeter initial mass
   function.
This is clearly demonstrated by the comparison of the SNe~IIP mass distribution,
   calculated by assuming a Salpeter initial mass function in the
   $9-25~M_{\sun}$ mass range, with the mass distribution of four SNe~IIP of
   known hydrodynamic mass (Fig.~\ref{fig:cumdis}).
Despite the small number of events, the significance of the difference between
   the two distributions is high: the probability that the four SNe occurred
   at random in the mass range of $15-25~M_{\sun}$ is only 0.01.
This indicates that either the 1D model of the SN explosion
   overestimates the ejecta mass, or the outcome of the core collapse of
   $9-15~M_{\sun}$ stars differs markedly from that of SNe~IIP observed until now.

To study the first possibility, we would require 3D radiation hydrodynamics modeling,
   which is not possible at present.
While we cannot readily ascertain any 3D effect that might reduce the required ejecta
   mass, apparent 3D effects do exist that could increase the hydrodynamic
   mass.
Indeed, the RT mixing between the helium core and the hydrogen envelope is
   expected to produce heterogeneous ejecta consisting of helium clumps
   embedded in the hydrogen background.
This structure should reduce unavoidably the average opacity of the hydrogen-rich
   matter.
As a result, the ejecta mass required to reproduce the observations should
   increase.
This could counterbalance other possible effects that might reduce the ejecta
   mass.
We believe therefore that the 3D hydrodynamic simulations are unlikely to reduce
   significantly the SN~IIP progenitor masses recovered from the 1D modeling.
The situation could be eventually clarified by 3D modeling of the SN~IIP
   explosion.

Alternatively, the disparity between the two distributions in
   Fig.~\ref{fig:cumdis} is real.
In this case two explanations could be invoked:
   (1) $9-15~M_{\sun}$ stars produce faint, still undetectable SNe~IIP,
   i.e. we have a selection effect;
   (2) core collapse of stars from this mass range does not produce an SN event
   at all.
In the latter case, the fate of the star may be silent collapse with a
   neutron star residing inside the stellar envelope, i.e. a Thorne-\.{Z}ytkow
   object (Thorne \& \.{Z}ytkow \cite{TZ_75}).
If this is the case, the neutron star may eventually grow into a black hole of
   mass as high as $\sim15~M_{\sun}$ due to the rapid ($\sim10^2$ years)
   accretion of the stellar envelope (Bisnovatyi-Kogan \& Lamzin \cite{BL_84}).

The second option appears, however, to be unlikely, because it implies that, apart
   from double neutron stars (DNS) originating from $9-25~M_{\sun}$ stars,
   there should be a comparable number of binaries with a neutron star in
   combination with a black hole (NSBH binaries).
Assuming that the production rate of DNS and NSBH binaries from
   $9-25~M_{\sun}$ stars is determined by the random pairing of stars with the
   Salpeter initial mass function and that $9-15~M_{\sun}$ stars produce black
   holes, we expect the relative rate of formation of these binaries to be
   $\mathrm{DNS}:\mathrm{NSBH}=1:0.85$ for $9-25~M_{\sun}$ stars.
This ratio is in an apparent contradiction with the fact that eight DNS in
   the Galaxy are known (Ihm et al. \cite{IKB_06} and references therein) and
   no NSBH binary has yet been discovered.
The probability of a random realization of this situation is only
   $1.5\times10^{-5}$, i.e. sufficiently small to be able to exclude the second option
   that stars in the mass range of $9-15~M_{\sun}$ end their lifes as black holes.

We propose that the core collapse of $9-15~M_{\sun}$ stars should
   produce a neutron star and the remaining stellar matter ejected as
   a result of a faint SN event.
This picture predicts that the rate of faint SNe~IIP should be comparable with
   the combined rate of normal (e.g., SN~1999em), sub-luminous (e.g., SN 2005cs),
   and low-luminosity (e.g., SN~2003Z) SNe~IIP.
A detection of the extended class of faint SNe~IIP, or non-detection at
   a low flux level, could verify this scenario for $9-15~M_{\sun}$ stars.
It is interesting that some known transient events, e.g. SN~1997bs
   (Van Dyk et al. \cite{VPK_00}), optical transient M85 OT2006-1
   (Pastorello et al. \cite{PDS_07}), and SN~2008S (Prieto et al. \cite{PKT_08})
   might belong to the proposed category of faint SNe related to the mass
   range of $9-15~M_{\sun}$.

\begin{acknowledgements}
VU is very grateful to Wolfgang Hillebrandt for the excellent opportunity
   to work at the MPA.
We thank the referee John Eldridge for careful reading of the manuscript and
   helpful comments.
\end{acknowledgements}


\end{document}